\documentclass[twoside]{ae100prg}
\usepackage{graphicx}
\usepackage[breaklinks]{hyperref}
\usepackage{booktabs}

\begin{document}
\title{Signature change in \\ loop quantum cosmology}

\author{Jakub Mielczarek}
\address{Theoretical Physics Department, National Centre for Nuclear Research,
 Ho{\.z}a 69, 00-681 Warsaw, Poland}
\email{jakubm@fuw.edu.pl}

\begin{abstract}
The Wick rotation is commonly considered only as an useful computational trick. However, as it 
was suggested by Hartle and Hawking  already in early eighties, Wick rotation may gain physical 
meaning at the Planck epoch.  While such possibility is conceptually interesting, leading to no-boundary 
proposal, mechanism behind the signature change remains mysterious. We show that the signature 
change anticipated by Hartle and Hawking naturally appear in loop quantum cosmology.
Theory of cosmological perturbations with the effects of quantum holonomies is discussed. It was 
shown by Cailleteau \textit{et al.} (Class. Quant. Grav. {\bf 29} (2012) 095010) that this theory can 
be uniquely formulated in the anomaly-free manner. The obtained algebra of effective constraints 
turns out to be modified such that the metric signature is changing from Lorentzian in low curvature 
regime to Euclidean in high curvature regime. Implications of this phenomenon on propagation 
of cosmological perturbations are discussed and corrections to inflationary power spectra of scalar 
and tensor perturbations are derived. Possible relations with other approaches to quantum gravity 
are outlined. We also propose an intuitive explanation of the observed signature change using
analogy with spontaneous symmetry breaking in ``wired" metamaterials. 
\end{abstract}

\section{Introduction}

The metric signature change from Lorentzian to Euclidean is usually performed by the so-called Wick rotation 
($t \rightarrow -i \tau $), under which the line element $ds^2 = -dt^2+d{\bf x}^2$ transforms to $ds^2 = d\tau^2
+d{\bf x}^2$. The Wick rotation becomes especially important in the path integral formulation of quantum 
mechanics. It allows to calculate non-perturbative effects by considering \textit{instantons}.  Another advantage 
is coming from the Wick rotation is improvement of a convergence property of some path integrals. But these are 
just useful computational tricks.     
 
However,  in 1983 Hartle and Hawking proposed that Wick rotation may gain physical meaning at the Planck 
epoch \cite{Hartle:1983ai}. This assumption was crucial for construction of the so-called no-boundary proposal, 
which was a way to cope with the problem of initial conditions for the Universe.  While such possibility is conceptually 
interesting, mechanism behind the signature change in the very early Universe remains enigmatic.

If such transition from the Lorentzian to Euclidean space has occurred in the early Universe, what could be 
the origin of this?  Can the signature change be due to some quantum gravity effects? So far, there were no 
indications supporting such possibility. However, recent results coming from symmetry reduced models of 
Loop Quantum Gravity (LQG)\cite{Ashtekar:2004eh} suggest that indeed the signature change may occur due 
to discrete nature of space at the Planck scale \cite{Cailleteau:2011kr,Bojowald:2011aa}. 

\section{Loops}

In LQG, granularity of space at the Planck scale is manifested by discrete spectra of geometric operators as 
area and volume. The starting point for constructing LQG is the Hamiltonian formulation of General Relativity 
in language of Ashtekar variables fulfilling $\left\{ E^a_j ({\bf x}),A^i_b({\bf y}) \right\} = 8\pi G \gamma 
\delta^a_b \delta^i_j \delta^{(3)}({\bf x-y})$, where $\gamma$ is a free parameter of the theory called 
Barbero-Immirzi parameter. In this framework, Hamiltonian of gravity sector can be written as a sum of  
three constraints: $H_{G}[N,N^a,N^i]  = S[N]+D[N^a]+G[N^i] \approx 0$. Here $S$ is the scalar constraint, 
$D$ is diffeomorphisms constraint and $G$ is the Gauss constraint. The constraints ($S \rightarrow \mathcal{C}_1, 
D \rightarrow \mathcal{C}_2, G \rightarrow \mathcal{C}_3 $) fulfill the closed algebra 
$\{ \mathcal{C}_I, \mathcal{C}_J \} = {f^K}_{IJ}(A^j_b,E^a_i) \mathcal{C}_K$,  
where ${f^K}_{IJ}(A^j_b,E^a_i) $ are some structure functions. 
 
Based on the Ashtekar variables, nonlocal variables called holonomies and fluxes are constructed. 
These new variables are subject of quantization in LQG.  For our purposes, it is sufficient to note that 
holonomy is defined as parallel transport of $A_a =A^i_a\tau_i$ along some curve $e$ on spatial 
hypersurface: $h_e := \mathcal{P} \exp \int_e A_a dx^a$, where $\sigma_j =2i \tau_j$ are Pauli matrices. 
The holonomies are elements of $SU(2)$ group.

\begin{figure}
\begin{center}
\begin{tabular}{ccc} 
\includegraphics[width=3cm]{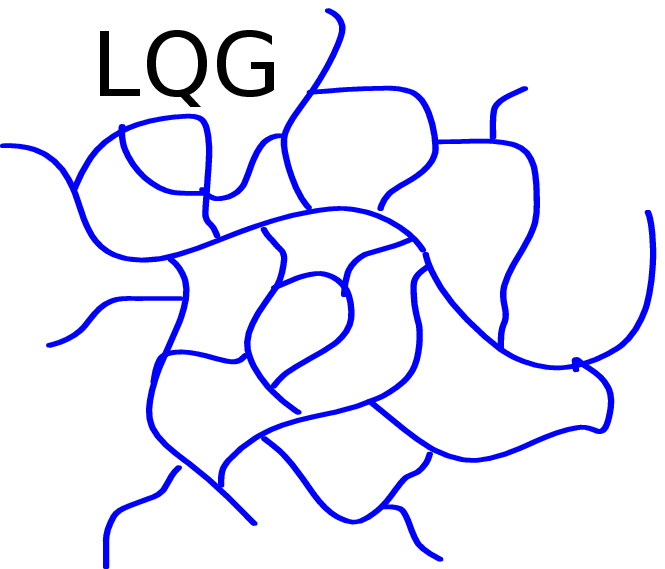} & \ \ \ \ \ \ \ & \includegraphics[width=3.5cm]{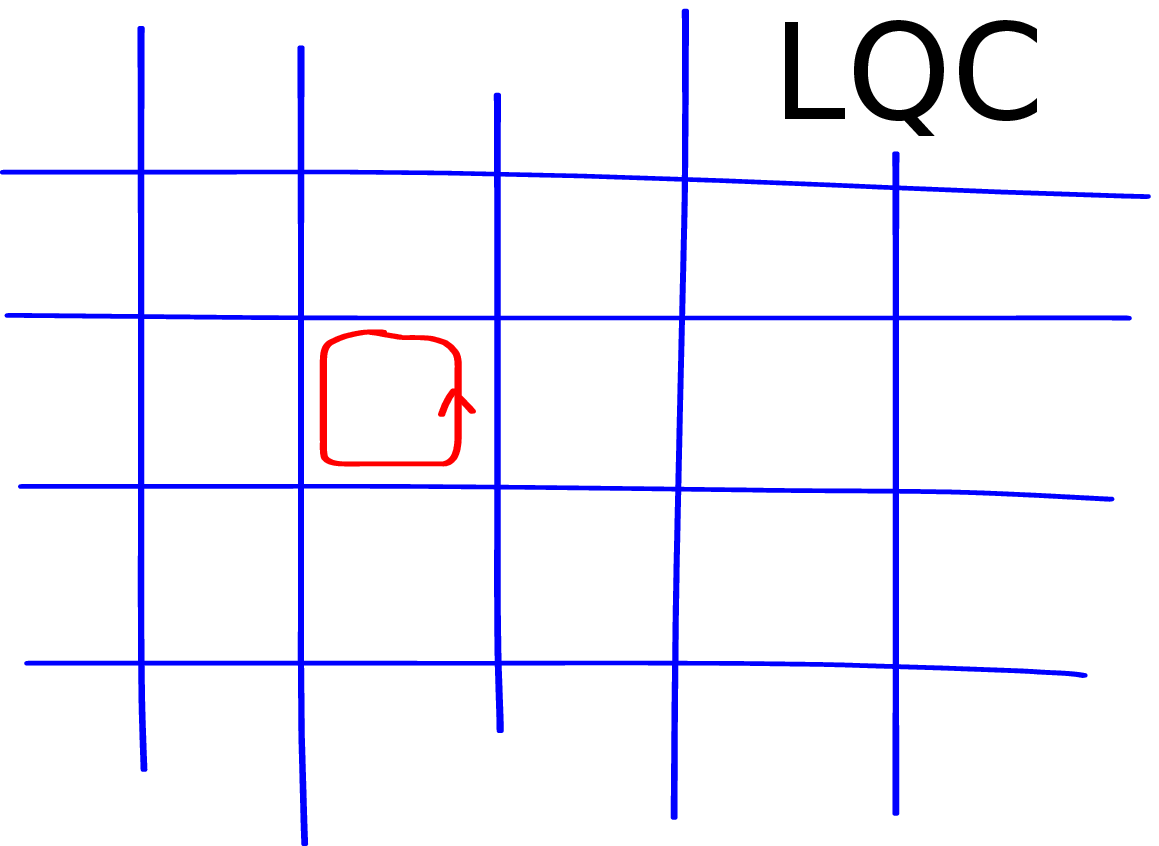}
\end{tabular}
\end{center}
\caption{\label{LQGLQC} In LQG, state of geometry is described by the spin network (left).  
In LQC, the spin network takes a form of regular lattice (right). The (red) loop represents 
holonomy $h_{\Box}$ around the elementary cell.}
\end{figure}

In LQG, a state of gravity is described by superposition of graphs called \emph{spin networks}. The 
links of the graphs are labelled by half integers ($j=1/2,1,3/2,\dots$) corresponding to irreducible 
representations of the $SU(2)$ group. An exemplary spin network is shown in Fig. \ref{LQGLQC}. 
 
Loop Quantum Cosmology (LQC)\cite{Bojowald:2006da} is a regular lattice model of LQG. In
particular, symmetries of isotropy and homogeneity are imposed on the spin network  within LQC. 
In what follows, we will consider isotropic model with small perturbative inhomogeneities around 
a flat Friedmann-Robertson-Walker (FRW) background.  

In LQC,  physical area of the elementary lattice cell Ar$_{\Box} = \bar{p} \bar{\mu}^2$, where 
$\bar{p}=a^2$ and $a$ is a scale factor. In general $\bar{\mu} \propto \bar{p}^{\beta}$, where
$-1/2 \leq \beta \leq 0$. For the so-called $\bar{\mu}-$scheme (``new quantization scheme"): 
$\bar{\mu}  = \sqrt{\frac{\Delta}{\bar{p}}}$, where usually $\Delta$ is assumed to be area gap 
derived from LQG: $\Delta = 2\sqrt{3} \pi \gamma l^2_{Pl}$. The $\bar{\mu}-$scheme, in which 
physical area of the elementary lattice cell is constant during the cosmological evolution, was 
shown to be physically favored.
  
At the effective level, effects of discreteness can be studied by introducing the so-called holonomy 
corrections. They are obtained by replacing curvature of Ashtekar connection by holonomy around 
elementary loop (see Fig. \ref{LQGLQC}).  This procedure is called \emph{polymerization}. 

\section{Cosmological perturbations}

In the most cosmological applications, Ashtekar variables can be decomposed into the background (here flat FRW) 
and perturbation parts: $E^a_i = \bar{E}^a_i  +\delta E^a_i$ and $A^i_a = \bar{A}^i_a +\delta A^i_a$, where 
$\bar{E}^a_i = \bar{p} \delta^a_i$ and $\bar{A}^i_a =\gamma \bar{k} \delta^i_a$. The perturbations of the Ashtekar 
variables can be related with the standard metric perturbations: scalar modes ($\Phi$, $\Psi$, E, B), vector modes 
($S^a$, $F_a$) and tensor modes ($h_{ab}$). In total, there is 10 perturbative degrees of freedom. Furthermore, 
matter degrees of freedom are also subject of perturbative decomposition. In what follows we consider model 
with a scalar field, so  $\varphi$ and its canonically conjugated momenta $\pi$ can be written as: $\varphi = \bar{\varphi}
+\delta \varphi$ and $\pi = \bar{\pi} +\delta \pi$. Applying the above decompositions, total constraints $\mathcal{C}_{tot}
=\mathcal{C}_{G}+\mathcal{C}_{M}$, which take into account contributions from gravity and matter, can be expanded. 
Our analysis is performed up to the second order in perturbative development: $\mathcal{C}_{tot} = \mathcal{C}^{(0)}
+\mathcal{C}^{(1)}+\mathcal{C}^{(2)}+...$, so the corresponding equations of motion stay linear. 

\section{Anomaly freedom and algebra of constraints}

The effects of discreteness of space are introduced by employing the holonomy corrections. Such corrections 
modify the classical constraints  $\mathcal{C}_{tot}$ to some new effective quantum constraints $\mathcal{C}^Q_{tot}$.  
The modification holds the correspondence principle, such that in the limit $\bar{\mu}\rightarrow 0$, the modified 
constraints $\mathcal{C}^Q_{tot} \rightarrow \mathcal{C}_{tot}$. The procedure of introducing quantum corrections 
suffers from various ambiguities. Moreover, the resulting algebra of modified constraints is in general not closed: 
\begin{equation}
\{ \mathcal{C}^Q_I, \mathcal{C}^Q_J \} = {g^K}_{IJ}(A^j_b,E^a_i) \mathcal{C}^Q_K+\mathcal{A}_{IJ}, 
\end{equation}
where $\mathcal{A}_{IJ}$ are some anomaly terms. Closure of algebra is required by mathematical consistency of 
the theory. So, the question is: Can we introduce quantum holonomy corrections in the anomaly-free manner 
(\textit{i.e.} such that $\mathcal{A}_{IJ} = 0$)? The answer turns out to be ``yes".  Moreover, there is a unique way 
of modifying constraints such that the algebra is closed \cite{Cailleteau:2011kr}. Additionally, the conditions of 
anomaly-freedom are fulfilled if and only if  $\beta =-1/2$, which corresponds to the $\bar{\mu}$-scheme. Therefore, 
the only remaining free parameter is the area of elementary lattice cell $\Delta$, which is however expected to be 
of the order or the Planck area $l^2_{Pl}$. 

The obtained algebra of the effective quantum constraint is \cite{Cailleteau:2011kr}:
\begin{eqnarray}
\left\{ D_{tot}[N^a_1],D_{tot} [N^a_2] \right\} &=& 0, \\
\left\{ S_{tot}[N],D_{tot}[N^a] \right\} &=& - S_{tot}[\delta N^a \partial_a \delta N],\\
\left\{ S_{tot}[N_1],S_{tot}[N_2] \right\} &=& \Omega D_{tot} \left[ \frac{\bar{N}}{\bar{p}} \partial^a(\delta N_2 - \delta N_1)\right].  
\label{StotStot} 
\end{eqnarray}
The algebra is closed but deformed with respect to the classical case due to presence of 
$\Omega$ in Eq. \ref{StotStot}. Therefore, general covariance is modified.  The new factor 
$\Omega$ can be expressed as follows: $\Omega =\cos(2\bar{\mu} \gamma\bar{k}) = 1 - 2\rho/\rho_c \in [-1,1]$, 
where $\rho$ is energy density of the scalar matter and the critical energy density    
$\rho_c  := \frac{3}{8\pi G \Delta} \sim \rho_{Pl}:= m_{Pl}^4$. What is the interpretation 
of the above deformation of the algebra of constraints? In order to answer this 
question let us recall the classical equivalent of the modified bracket (\ref{StotStot}) for space with 
signature $\sigma$ \cite{Ashtekar:2004eh}: 
\begin{equation}
\left\{ S_{tot}[N_1],S_{tot}[N_2] \right\} = \sigma D\left[  \frac{\bar{N}}{\bar{p}} \partial^a(\delta N_2 - \delta N_1)\right].  
\nonumber
\end{equation}
Here, $\sigma =1$ corresponds to the Lorentzian signature and $\sigma =-1$ to the Euclidean one.
Therefore, we conclude that modification of the effective algebra of constraints (\ref{StotStot}) 
means that space becomes Euclidian for $\rho > \rho_c/2$, while Lorentzian geometry emerges for
$\rho < \rho_c/2$. In the regime of high curvatures and hight energy densities ($\rho > \rho_c/2$), 
spacetime becomes 4-dimensional Euclidean space.  There is no distinguished time direction in this
phase.  It is interesting to notice that this model exhibits properties of the Hartle-Hawking no-boundary 
proposal.  However here, signature change occurs smoothy with increase of energy density. 

The same effect of signature change was observed also for inhomogeneous spherically symmetric models 
with holonomy corrections \cite{Bojowald:2011aa}. Therefore, we have good reason to believe that this
phenomenon is a general consequence of quantum polymerization of space. Moreover, we can 
speculate that the \textit{of-shell} algebra of quantum constraints in LQG should also exhibit such 
property, \textit{i.e.} $[ \hat{S},\hat{S}] =  i \Omega  \hat{D}$. 

Among many other comments, the fact that \emph{ultralocal gravity} \cite{Isham:1975ur},
where $\{S,S\}=0$, is recovered at the transition point $\rho =\rho_{c}/2$ is worth stressing. 

\section{Equations of motion}

The obtained anomaly-free formulation can be now used to derive equations of motion 
for both background variables and perturbations. The background dynamics is governed 
by the modified Friedmann equation $H^2 = \frac{8\pi G}{3} \rho  \left(1-\rho/\rho_c \right)$,
where $H$ is the Hubble factor. Clearly, only $\rho \leq \rho_c$ are physically allowed, 
which was used to determine the range of $\Omega$. 

For the scalar perturbations one can define gauge-invariant variable $v$ and the 
corresponding modified Mukhanov equation \cite{Cailleteau:2011kr}:
\begin{eqnarray}
\frac{d^2}{d\tau^2}{v}-\Omega \nabla^2 v - \frac{z^{''}}{z} v = 0,
\end{eqnarray}
where $z := \sqrt{\bar{p}} \frac{\dot{\varphi}}{H} $. Here, $\tau = \int dt/a$ is a conformal time. 
For the considered model with a scalar field, vector modes are pure gauge, therefore do not 
contribute \cite{Mielczarek:2011ph}. Equation for tensor modes takes the form \cite{Cailleteau:2012fy}
\begin{equation}
\frac{d^2}{d\tau^2}h_{ab}+2\left( a H-\frac{1}{2\Omega} \frac{d\Omega}{d\tau}\right)\frac{d}{d\tau}h_{ab} 
-\Omega\nabla^2h_{ab} = 0. \nonumber
\end{equation}
The obtained equations are modified by presence of $\Omega$ in front of the Laplace operator. 
Therefore, transition to the Euclidean domain leads to change of equation type from 
hyperbolic to elliptic, as expected. Furthermore, we see that the speed of propagation is varying, 
since $c_s^2 = \Omega$.

\section{Holonomy corrections to inflationary power spectra}

As an application of the obtained equations of motion we will derive holonomy corrections to the 
inflationary scalar and tensor power spectra. We will focus on the slow-roll inflationary model driven 
by a single scalar field $\varphi$ with potential $V(\varphi)$ occurring in the Lorentzian domain.
The slow-roll parameters with the holonomy corrections are: 
\begin{eqnarray}
\epsilon :=  \frac{m_{Pl}^2}{16 \pi} \left( \frac{V_{,\varphi}}{V} \right)^2\frac{1}{(1-V/\rho_c)}
\ \ \textrm{and} \ \ \eta :=  \frac{m_{Pl}^2}{8 \pi} \left( \frac{V_{,\varphi\varphi}}{V} \right)  \frac{1}{(1-V/\rho_c)}. 
\nonumber
\end{eqnarray}

Derivation of the scalar and tensor power spectra is based on application of the standard techniques 
of the quantum field theory on curved spaces. Moreover, normalization is such that in the UV limit the 
Minkowski vacuum is recovered. Obtained spectra of scalar and tensor (gravitational waves)  perturbations are
\begin{equation}
\mathcal{P}_{S}(k) = A_S\left(\frac{k}{aH}\right)^{n_{S}-1} \ \   \textrm{and} \ \ 
\mathcal{P}_{T}(k) = A_T \left(\frac{k}{aH}\right)^{n_{T}}, \nonumber
\end{equation}
where amplitudes and spectral indices are given as follows: 
\begin{eqnarray}
A_S &=& \frac{1}{\pi \epsilon} \left(\frac{H}{m_{Pl}} \right)^2  \left(1+2 \frac{V}{\rho_c}  \right) 
\ \ \textrm{and} \ \ n_{S} =1+2\eta-6\epsilon \left(1-\frac{V}{\rho_c} \right), \nonumber \\
A_T &=& \frac{16}{\pi} \left(\frac{H}{m_{Pl}} \right)^2  \left(1+3 \frac{V}{\rho_c}  \right) 
\ \ \textrm{and} \ \ n_{T} =-2\epsilon\left(1-3 \frac{V}{\rho_c}\right). \nonumber
\end{eqnarray}
Furthermore, the consistency relation 
\begin{equation}
r := \frac{A_T}{A_S} \simeq16\epsilon  \left(1+\frac{V}{\rho_c}  \right). 
\end{equation}
The corrections are introduced by the factors $V/\rho_c$, which are of the order of $10^{-12}$ 
for typical values of parameters. Confrontation of the obtained spectra with the available 
CMB data will be studied elsewhere \cite{MKS2012_CMB}.  In more detailed analysis, initial 
conditions should be established at the transition point $\rho=\rho_c/2$, what can lead to some 
additional modification of the power spectra. This issue will be investigated in our further research. 

\section{Towards understanding the signature change}

It is tempting to understand origin of the signature change at a microscopic level.
Is this a kind of phase transition occurring at the level of the spin network 
or, in more general, at the level of some four dimension spin foam model? One possibility 
is that some sort of spontaneous symmetry breaking takes place, leading to distinction
of the time dimension in low curvature regime. 

A relevant example of spontaneous symmetry breaking is given by ferromagnets.  At hight 
temperatures, ferromagnets are loosing their magnetic properties. The spins (magnetic moments) 
are randomly orientated, and no direction is distinguished. The system satisfies the $SO(3)$ 
rotational symmetry, which is also a symmetry of the corresponding Hamiltonian. However, 
while temperature is decreased, spins starts to orientate is some direction and magnetic 
domains are formed. This process begins when temperature is lowered below the so-called 
Curie temperature. 

One can speculate that an analogous phase transition occurs in case of gravity. Namely, at high energies, 
the symmetry is, let say $SO(4)$. There is no distinction of any time coordinate. However, while lowering 
the energy density, which is an analogue of temperature, the symmetry will be broken to $SO(3)$. 
The energy density $\rho_c/2$ is an analogue of the Curie temperature. It is possible that $SO(3)$ 
symmetry of the triad rotations in Ashtekar formalism (doubly covered by $SU(2)$ group in LQG) 
is in fact a residual symmetry of some wider symmetry before spontaneous symmetry breaking. 

\begin{figure}
\begin{center}
\includegraphics[width=7cm]{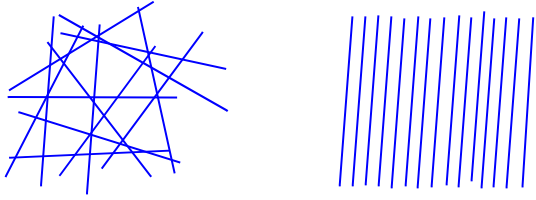}
\end{center}
\caption{\label{Wires} An hight temperatures orientation of nanowires is random (left). 
At low temperatures, some direction is distinguished due to the spontaneous 
symmetry breaking (right). In the distinguished direction, dielectric permittivity is 
negative, leading to emergence of a time variable.}
\end{figure}

The phase transition, similar to the one discussed in case of ferromagnets, may occur for the so-called  
``wired" metamaterials composed of nanowires. It was shown that for such materials, an effective 
emergence of time variable may occur because of \emph{negative dielectric permittivity} \cite{metamaterials}.  
At high temperatures, dielectric permittivities in all directions are positive. However in low temperature state, 
nanowires may align in some direction as spins do. In this distinguished direction, dielectric permittivity becomes 
negative leading to emergence of "time" direction at the level of equations of motion for electromagnetic 
field (see Fig. \ref{Wires}).        
We speculate that, the same kind of process occurs in case of gravity, leading to emergence of a
time variable.  Structure of four dimensional Euclidian space undergoes phase transition, 
such that some particular direction is picked (presumably domains with different time 
directions can form). Equations of motion for the fields living on such frame change from 
elliptic to hyperbolic, which is interpreted as emergence of the time direction.  We will explore 
this interpretation in more details in our further studies.  

\section{Summary and outlook}

We have shown that metric signature change may occur due to polymerization of space at the 
Planck scale.  Preliminary analysis of this new phenomenon was carried our.  Many questions 
remain open and are awaiting detailed analysis. In particular: Is there signature change in full 
LQG also? What are the initial conditions at $\rho=\rho_c/2$? What is happening at the microscopic 
scale? How is the propagation of hight energy photons affected? And many, many others. 

Summarizing, the paradigm shift seems to be observed in LQC. There is no longer deterministic 
bouncing phase as was thought for many years. The Big Bounce model in LQC seems to be an 
artifact of the strong assumption of homogeneity.  Due to the Euclidean stage there is no 
access to the information contained in contacting Universe, which gives answer to the long 
standing debate on cosmic forgetfulness. 
 
\section*{Acknowldegments}

I would like thank to Aur\'elien Barrau, Thomas Cailleteau and Julien Grain 
for excellent collaboration and W{\l}odzimierz Piechocki for his careful reading 
of the manuscript and useful remarks.

\end{document}